\documentclass[twocolumn]{jpsj3}

\def\runauthor{M. Koga, M. Matsumoto, and H. Kusunose}

\title{
Effects of Impurities with Singlet-Triplet Configuration on Multiband Superconductors
}

\author{Mikito Koga, Masashige Matsumoto$^1$, and Hiroaki Kusunose$^2$}

\inst{
Department of Physics, Faculty of Education, Shizuoka University, Shizuoka
422--8529, Japan \\
$^1$Department of Physics, Faculty of Science, Shizuoka University, Shizuoka
422--8529, Japan \\
$^2$Department of Physics, Ehime University, Matsuyama 790-8577, Japan
}

\recdate{\today}

\abst{ 
Roles of multipole degrees of freedom in multiband superconductors are investigated
in a case of impurities whose low-lying states consist of singlet ground and triplet excited states,
which is related to the experimental fact that the transition temperature $T_{\rm c}$ is increased
by Pr substitution for La in LaOs$_4$Sb$_{12}$.
The most important contribution to the $T_{\rm c}$ increase comes from the inelastic interband
scattering of electrons coupled to quadrupole or octupole moments of impurities.
It is found that a magnetic field modifies an effective pairing interaction and the scattering
anisotropy appears in the field-orientation dependence of the upper critical field $H_{{\rm c}2}$
in the vicinity of $T_{\rm c}$, although a uniaxial anisotropic field is required for experimental
detection.
This would be proof that the Pr internal degrees of freedom are relevant to the stability of
superconductivity in (La$_{1-x}$Pr$_x$)Os$_4$Sb$_{12}$.
}

\kword{
heavy fermion, skutterudite, impurity, crystal field, multiband superconductivity
}

\begin{document}

\maketitle

%%%%%%%%%%%%%%%%%%%%%%%%%%%%%%%%%%%%%%%%%%%%%%%%%%%%%%%%%%%%%%%%%%%%%%%%%%%%%%%%
%Macros
%%%%%%%%%%%%%%%%%%%%%%%%%%%%%%%%%%%%%%%%%%%%%%%%%%%%%%%%%%%%%%%%%%%%%%%%%%%%%%%%
\newcommand{\ds}{\displaystyle}

\renewcommand{\H}{\mathcal{H}}
\newcommand{\br}{{\mbox{\boldmath$r$}}}
\newcommand{\bR}{{\mbox{\boldmath$R$}}}
\newcommand{\bS}{{\mbox{\boldmath$S$}}}
\newcommand{\bk}{{\mbox{\boldmath$k$}}}
\newcommand{\bH}{{\mbox{\boldmath$H$}}}
\newcommand{\bh}{{\mbox{\boldmath$h$}}}
\newcommand{\bJ}{{\mbox{\boldmath$J$}}}
\newcommand{\bPsi}{{\mbox{\boldmath$\Psi$}}}
\newcommand{\bpsi}{{\mbox{\boldmath$\psi$}}}
\newcommand{\bPhi}{{\mbox{\boldmath$\Phi$}}}
\newcommand{\bG}{{\hat{G}}}
\newcommand{\om}{{\omega_l}}
\newcommand{\omd}{{\omega^2_l}}
\newcommand{\btau}{{\hat{\tau}}}
\newcommand{\brho}{{\hat{\rho}}}
\newcommand{\bsigma}{{\hat{\sigma}}}
\newcommand{\bSigma}{{\hat{\Sigma}}}
\newcommand{\bI}{{\hat{I}}}
\newcommand{\bt}{{\hat{t}}}
\newcommand{\bq}{{\hat{q}}}
\newcommand{\bLambda}{{\hat{\Lambda}}}
\newcommand{\bDelta}{{\hat{\Delta}}}
\newcommand{\bskp}{{\mbox{\scriptsize\boldmath $k$}}}
\newcommand{\skp}{{\mbox{\scriptsize $k$}}}
\newcommand{\bsrp}{{\mbox{\scriptsize\boldmath $r$}}}
\newcommand{\bsRp}{{\mbox{\scriptsize\boldmath $R$}}}
\newcommand{\bsk}{\bskp}
\newcommand{\sk}{\skp}
\newcommand{\bsr}{\bsrp}
\newcommand{\bsR}{\bsRp}
\newcommand{\ri}{{\rm i}}
\newcommand{\re}{{\rm e}}
\newcommand{\rd}{{\rm d}}
\newcommand{\Tc}{{$T_{\rm c}$}}
\renewcommand{\Pr}{{PrOs$_4$Sb$_{12}$}}
\newcommand{\La}{{LaOs$_4$Sb$_{12}$}}
\newcommand{\LaPr}{{(La$_{1-x}$Pr${_x}$)Os$_4$Sb$_{12}$}}
\newcommand{\PrLa}{{(Pr$_{1-x}$La${_x}$)Os$_4$Sb$_{12}$}}
\newcommand{\OsRu}{{Pr(Os$_{1-x}$Ru$_x$)$_4$Sb$_{12}$}}
\newcommand{\PrRu}{{PrRu$_4$Sb$_{12}$}}
%%%%%%%%%%%%%%%%%%%%%%%%%%%%%%%%%%%%%%%%%%%%%%%%%%%%%%%%%%%%%%%%%%%%%%%%%%%%%%%%%%%%%%%%%%%%%%%%%%%

%%%%%%%%%%%%%%%%%%%%%%%%%%%%%%%%%%%%%%%%%%%%%%%%%%%%%%%%%%%%%%%%%%%%%%%%%%%%%%%%%%%%%%%%%%%%%%%%%%%
\section{Introduction}
%%%%%%%%%%%%%%%%%%%%%%%%%%%%%%%%%%%%%%%%%%%%%%%%%%%%%%%%%%%%%%%%%%%%%%%%%%%%%%%%%%%%%%%%%%%%%%%%%%%
It is common knowledge that magnetic impurities induce pair breaking and a decrease in
superconducting transition temperature \Tc.
In a multiband case, however, there is room to reconsider such a conventional understanding.
\cite{Golubov97,Arseev02}
Electron scattering by impurities occurs not only within one band (intraband) but also between
different bands (interband).
In addition, variations in order parameters can be considered for superconducting bands.
In a two-band case, for instance, the $s_\pm$-wave state is
characterized by the sign-reversing order parameters.
\cite{Mazin08,Kuroki08}
According to recent studies on the $s_\pm$-wave state, interband scattering does not contribute
to \Tc~suppression for magnetic impurities.
\cite{Matsu09,Li09}
This behavior resembles that of intraband scattering by nonmagnetic impurities in a single-band
$s$-wave superconductor.
Furthermore, \Tc~can be increased by inelastic interband scattering if the impurities
have such internal degrees of freedom as crystal-field split energy levels.
\cite{Koga10}
The idea was originally proposed by Fulde {\it et al.} for nonmagnetic impurities
in a single-band superconductor;
\cite{Fulde70}
however, it has been considered that magnetic impurities always induce \Tc~suppression.
We have to check carefully the potential roles of impurities with orbital degrees of freedom,
e.g., multipoles in $f$-electron systems.
\par

Since the discovery of the heavy fermion superconductor \Pr~($T_{\rm c} = 1.85$ K),
\cite{Bauer02}
much attention has been attracted by the following unique superconducting properties:
\cite{Aoki07}
(1)~Change in gap symmetry in a magnetic field.
This was first observed as the field orientation dependence of oscillating patterns on
thermal conductivity,
\cite{Izawa03}
although no symptom of cubic symmetry breaking was found
in a recent angle-resolved specific heat experiment.
\cite{Custers06}
(2)~Broken time reversal symmetry below \Tc.
An internal magnetic field emerges spontaneously but is extremely small, which was detected by
muon spin relaxation measurement.
\cite{Aoki03}
The identification of the pairing parity is under debate.
(3)~Robustness against substitution effects.
Substituting La for Pr leads to the gradual decrease in \Tc.
\cite{Rotundu06}
Surprisingly, the $x$ dependence of \Tc~in \PrLa~is smoothly connected to the \Tc($=0.74$ K) of
\La~that exhibits the conventional $s$-wave property.
\cite{Rotundu06,Yogi06}
In another experiment on \OsRu, \Tc~gradually decreases until $x \sim 0.6$ and then shows
a monotonic increase up to the \Tc($=1.3$ K) of \PrRu, which is also an $s$-wave superconductor.
\cite{Frederick04}
It has been considered that these unconventional properties are associated with
the Pr $4f$-electron behavior, although this association is still mysterious.
\par

It is also important to note that \Pr~exhibits the quadrupole ordering in a higher magnetic field
where the superconductivity disappears.
\cite{Aoki07,Aoki02,Kohgi03}
This ordering phase diagram is successfully explained by the localized Pr $4f^2$
quasi-quartet model.
\cite{Shiina04a,Shiina04b}
In fact, it has been established by the recent experiments that the Pr low-lying states consist of
$\Gamma_1$ singlet ground and $\Gamma_4^{(2)}$ excited triplet states well separated from the
higher crystal-field energy levels.
\cite{Kuwahara05,Tou05}
The $\Gamma_4^{(2)}$ representation is for the $T_h$ point group with no fourfold symmetry axis.
\cite{Takegahara01}
This wave function is expressed by a combination of $\Gamma_4$ and $\Gamma_5$ wave
functions in the $O_h$ point group, where the latter $\Gamma_5$ is more dominant.
Owing to the small singlet-triplet energy splitting $\simeq 8$ K,
it is expected that the quadrupole degrees of freedom will play a key role in the unconventional
superconductivity as well as in the field-induced ordering.
However, there is no reason to deny any contribution from other multipole degrees of freedom
in the Pr singlet-triplet configuration.
\par

To find a clue to understand the multipole contribution to the superconductivity,
we focus on the Pr impurity effect on the \La~superconductor.
As mentioned above, \Tc~is increased by Pr substitution for La in \La.
In addition, multiband properties are indicated by thermal-transport measurement under
a magnetic field in \Pr
\cite{Seyfarth05}
and by nuclear quadrupole resonance measurement in \LaPr.
\cite{Yogi06}
Suppose that \La~is a single-band superconductor and that the inelastic scattering by the Pr
impurities contributes to the \Tc~increase, the most probable origin of the \Tc~increase is a
quadrupolar scattering effect that leads to an effective pairing interaction that stabilizes the
superconductivity, by analogy with the optical phonon-mediated pairing.
\cite{Chang07}
If a multiband picture is applicable, there are two possibilities:
For the $s_{++}$-wave state with the same sign order parameters,
interband nonmagnetic (quadrupolar) scattering is also favorable for the \Tc~increase by the
inelastic impurity scattering, while interband magnetic (octupolar) scattering can increase \Tc~for
the $s_\pm$-wave state.
The relevance of the multiband scenario is closely connected to the characteristic structure of
skutterudites.
In \Pr, each Pr ion is located at the center of the Sb$_{12}$ icosahedron cage.
The most important point is the local hybridization of the Pr $4f$-electron states with conduction
bands via the Sb$_{12}$ molecular orbitals denoted by the $a_u$ and $t_u$ point-group symmetries.
\cite{Harima03,Otsuki05}
If the $a_u$-$t_u$ orbital exchange is the most relevant for the Pr impurity scattering,
the multipolar coupling in the Pr singlet-triplet configuration contributes to the \Tc~increase.
In our previous work, we proposed a possibility of \Tc~increase by magnetic impurity
scattering in the $s_{++}$-wave state erroneously instead of the $s_\pm$-wave state.
\cite{Koga10,Koga11}
We will give a correct description of the impurity scattering effect on the two-band
superconductivity in this case.
\par

In this paper, we discuss the multiorbital scattering effect on the multiband superconductivity that
reflects the local orbital symmetry.
In the Pr singlet-triplet configuration, electrons are coupled to the quadrupoles expressed as
$yz$, $zx$, and $xy$ or the octupoles expressed as $x(y^2 - z^2)$, $y(z^2 - x^2)$, and
$z(x^2 - y^2)$ by analogy with the dipoles expressed as $x$, $y$, and $z$, respectively.
\cite{Kuramoto09}
We focus on the magnetic octupolar scattering effect throughout the paper.
There is a marked distinction between such multipoles and the spin when a magnetic field is
introduced.
For the orbital exchange scattering by a multipole moment, we find that its polarization
changes as the field direction is rotated, while the spin exchange scattering is isotropic.
If such orbital exchange scattering is an origin of \Tc~increase, an effective pairing interaction
is modified by the scattering anisotropy under the field.
It is expected that the anisotropy effect will be observed as the field orientation dependence of
an upper critical field $H_{{\rm c}2} (T)$ line.
This would be  conclusive evidence indicating that the orbital degrees of freedom definitely play a
crucial role in the \Tc~increase, which is closely connected to the unique Pr atomic structure
in \LaPr.
\par

The paper is organized as follows.
In \S2, we show a typical form of the inelastic magnetic scattering that couples the
$O_h$ $\Gamma_1$ singlet and $\Gamma_5$ triplet states.
In \S3, we explain a gap equation for \Tc~increased or reduced by interband magnetic
scattering as an impurity effect on the two-band $s$-wave superconductivity.
The same formulation is applied straightforwardly to an interband nonmagnetic scattering case.
The gap equation is modified by including a magnetic field effect on impurities.
In \S4, this argument is extended to a case of single $a_u$ and threefold degenerate $t_u$
bands that reflects the cubic symmetry.
Owing to the Zeeman splitting of the excited triplet, \Tc~depends on the field direction.
Since the \Tc~deviation is very small when the cubic symmetry is conserved,
we consider a uniaxial anisotropy effect in the gap equation to elucidate the close correlation
between the field-orientation-dependent \Tc~and the anisotropy of multipolar scattering.
Finally, conclusions are given in \S5.

%%%%%%%%%%%%%%%%%%%%%%%%%%%%%%%%%%%%%%%%%%%%%%%%%%%%%%%%%%%%%%%%%%%%%%%%%%%%%%%%%%%%%%%%%%%%%%%%%%%
\section{Inelastic Scattering by Impurities}
%%%%%%%%%%%%%%%%%%%%%%%%%%%%%%%%%%%%%%%%%%%%%%%%%%%%%%%%%%%%%%%%%%%%%%%%%%%%%%%%%%%%%%%%%%%%%%%%%%%
First, we show a typical case of the inelastic electron scattering by magnetic impurities, keeping in
mind the Pr$^{3+}$ states in \Pr.
We consider that the low-lying states consist of the $\Gamma_1$ singlet ground and
$\Gamma_5$ triplet excited states in an $O_h$ crystal field.
They are expressed by
\cite{Kuramoto09}
\begin{align}
& | \Gamma_1 \rangle = \frac{\sqrt{30}}{12} (| 4 \rangle + | -4 \rangle ) + \frac{\sqrt{21}}{6} | 0 \rangle, \\
& \left\{
\begin{array}{l}
| \Gamma_5 + \rangle = \sqrt{\ds{\frac{7}{8}}} | 3 \rangle - \sqrt{\ds{\frac{1}{8}}} | -1 \rangle, \\
| \Gamma_5 0 \rangle = \sqrt{\ds{\frac{1}{2}}} (| 2 \rangle - | -2 \rangle ), \\
| \Gamma_5 - \rangle = -\sqrt{\ds{\frac{7}{8}}} | -3 \rangle + \sqrt{\ds{\frac{1}{8}}} | 1 \rangle,
\end{array}
\right.
\label{eqn:Gamma5}
\end{align}
where $| M \rangle$ ($M = 4, 3, \cdots , -4$) is an eigenstate of $J_z$ for the $J = 4$ total angular
momentum.
The interchange of the singlet and triplet states is caused by the local orbital
exchange of electrons via the hybridization of $f$-orbitals with conduction bands.
Since each Pr ion is located at the center of the Sb$_{12}$ cage, the Sb$_{12}$ molecular
orbitals mediate the hybridization, which is the most pronounced feature of skutterudites.
The $a_u$ ($\Gamma_2$) molecular orbital mostly contributes to the main conduction band
($a_u$ band), hybridizing with the $f$-orbitals most strongly.
\cite{Harima03}
The spin and orbitally coupled states of $f$-electrons are categorized  as the
$\Gamma_7$ representation.
We also consider here that the $t_u$ ($\Gamma_4$) orbitals participate in the secondary
conduction band ($t_u$ band), having a weaker hybridization with the $\Gamma_8$
$f$-electron states (see Appendix~A).
\par

%%%%%%%%%%%%%%%%%%%%%%%%%%%%%%%%%%%%%%
\begin{figure}
\begin{center}
\includegraphics[width=7cm,clip]{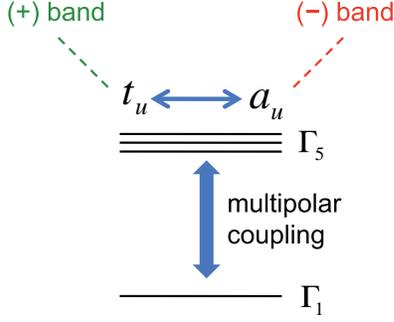}
\end{center}
\caption{Sketch of the inelastic multipolar exchange scattering with the $a_u$-$t_u$
orbital exchange in the Pr singlet-triplet configuration.
Here, the local $t_u$ and $a_u$ electrons participate in the different bands denoted by ($+$) and
($-$), respectively.
}
\label{fig:1}
\end{figure}
%%%%%%%%%%%%%%%%%%%%%%%%%%%%%%%%%%%%%%

Both quadrupole and octupole moments with the $\Gamma_5$ symmetry are involved in the
interchange of the Pr $\Gamma_1$ and $\Gamma_5$ states.
We focus on the inelastic octupolar scattering since it is unconventional that
\Tc~ can be increased by magnetic correlations.
In the present case, the octupolar exchange scattering is accompanied by the $a_u$-$t_u$ orbital
exchange of local electrons, which is shown in Fig.~\ref{fig:1}.
This effective exchange interaction is expressed by the following Hamiltonian
$\H_{\rm ex} = \H_{\rm I} + \H'$:
\begin{align}
& \H_{\rm I} = \sum_{\bsR_\gamma} \sum_{n} \delta_n a_{\gamma n}^\dagger a_{\gamma n},
\label{eqn:H-I} \\
& \H' = J_S \sum_{\bsR_\gamma} \sum_{nn'} \int\rd\br
        a_{\gamma n}^\dagger a_{\gamma n'} \delta(\br-\bR_\gamma) \cr
&~~~~~~~~~~~~~~~~~~\times
        \bpsi^\dagger(\br) \left( \bI_S \right)_{nn'} \bpsi (\br).
\label{eqn:H'}
\end{align}
The first term $\H_{\rm I}$ is for the impurity states, where $\bR_\gamma$ represents the position
of the $\gamma$th impurity and $a_{\gamma n}^\dagger$ ($a_{\gamma n}$) is the
pseudo-fermion creation (annihilation) operator for the $n$th impurity energy level $\delta_n$ at
the $\gamma$th impurity site ($n = 1, 2, 3$, and $4$ denote $\Gamma_1$, $\Gamma_5 +$,
$\Gamma_5 0$, and $\Gamma_5 -$, respectively).
\cite{Abrikosov65}
In the second term $\H'$, $\bI_S$ represents the $\Gamma_5$-type octupolar exchange
scattering of conduction electrons with the coupling constant $J_S$ 
($S$ denotes the magnetic scattering here) that is accompanied by the interchange among
the $n$th and $n'$th energy levels:
\begin{align}
\left( \bI_S \right)_{nn'} = \left( T_z \right)_{nn'} \bt_z + \frac{1}{2} \left( T_+ \right)_{nn'} \bt_-
                                           + \frac{1}{2} \left( T_- \right)_{nn'} \bt_+,
\label{eqn:I-S1}
\end{align}
where $T_\eta$ and $\bt_\eta$ ($\eta = z, \pm$) are octupole operators for the impurity states
and conduction electrons, respectively.
Their complete expressions are given in \S3 of ref.~31.
We use the two-band expression $ \bpsi^\dagger \bt_\eta \bpsi$
for the octupolar exchange scattering with the $a_u$-$t_u$ orbital exchange as described in
Appendix~A,  where we have corrected the previous results.
\cite{Koga10}

%%%%%%%%%%%%%%%%%%%%%%%%%%%%%%%%%%%%%%%%%%%%%%%%%%%%%%%%%%%%%%%%%%%%%%%%%%%%%%%%%%%%%%%%%%%%%%%%%%%
\section{Gap Equation for Two-Band Superconductivity}
%%%%%%%%%%%%%%%%%%%%%%%%%%%%%%%%%%%%%%%%%%%%%%%%%%%%%%%%%%%%%%%%%%%%%%%%%%%%%%%%%%%%%%%%%%%%%%%%%%%
Before discussing the multipolar scattering effect on \Tc, we briefly review our previous
study and give a correct description for the the Pr-like impurities with the singlet-triplet
configuration, focusing on how to derive a gap equation for the two-band
$s_{++}$-wave and $s_\pm$-wave superconducting states.
\cite{Koga10}
A magnetic field effect, which modifies an effective pairing interaction, is also taken into account
in the gap equation.
In Table~\ref{table:1}, we summarize the effective interaction type, which is either attractive
or repulsive, mediated by the interband magnetic (octupolar) scattering or nonmagnetic
(quadrupolar) scattering impurities in each $s$-wave state.
\par

%%%%%%%%%%%%%%%%%%%%%%%%%%%%%%%%%%%%%%
\begin{table}
\caption{Effective pairing interaction type mediated by the interband magnetic or nonmagnetic
(octupolar or quadrupolar, respectively, in the $O_h$ $\Gamma_1$-$\Gamma_5$ configuration)
scattering impurities: A (attractive) or R (repulsive).
It depends on the pairing type used.
}
\label{table:1}
\begin{center}
\begin{tabular}{ccc}
\hline
Scattering type & Pairing type &  Interaction type \\
\hline
magnetic & $s_{++}$-wave & R \\
~ & $s_\pm$-wave & A \\
\hline
nonmagnetic & $s_{++}$-wave & A \\
~ & $s_\pm$-wave & R \\
\hline
\end{tabular}
\end{center}
\end{table}
%%%%%%%%%%%%%%%%%%%%%%%%%%%%%%%%%%%%%%

%%%%%%%%%%%%%%%%%%%%%%%%%%%%%%%%%%%%%%%%%%%%%%%%%%%%%%%%%%%%%%%%%%%%%%%%%%%%%%%%%%%%%%%%%%%%%%%%%%%
\subsection{Formulation}
%%%%%%%%%%%%%%%%%%%%%%%%%%%%%%%%%%%%%%%%%%%%%%%%%%%%%%%%%%%%%%%%%%%%%%%%%%%%%%%%%%%%%%%%%%%%%%%%%%%
To describe the superconductivity, we start from the following Hamiltonian for the two-band
($\mu = \pm$) electrons:
\begin{align}
& \H_{\rm C} = \sum_{\mu \sigma} \int\rd\br \psi_{\mu \sigma}^\dagger(\br) \epsilon(-\ri\nabla)
\psi_{\mu\sigma}(\br)
\cr
&~~~~~~ - \sum_\mu \Delta_\mu \int\rd\br
\left[ \psi_{\mu \uparrow}^\dagger(\br)\psi_{\mu \downarrow}^\dagger(\br)
+ \psi_{\mu \downarrow}(\br)\psi_{\mu \uparrow}(\br) \right]. \cr
&
\label{eqn:H-C}
\end{align}
Here, we assume that  both bands are identical and the order parameters have the same
amplitude $|\Delta_+| = |\Delta_-| = \Delta$ for simplicity.
The operator $ \epsilon(-\ri\nabla) = - \nabla^2 / (2m_{\rm e}) - E_{\rm F}$ expresses the kinetic
energy measured from the Fermi energy $E_{\rm F}$, where $m_{\rm e}$ is the electron mass
and $\hbar = 1$.
It is convenient to introduce the $8\times 8$ matrix form of the thermal Green's function
\begin{align}
\bG(\tau,\br,\br') = - \langle T \bPsi(\br,\tau) \bPsi^\dagger(\br',0) \rangle,
\end{align}
with the eight-dimensional vectors $\bPsi(\br)$ and $\bPsi^\dagger(\br)$ for the two-band
electrons defined as
\begin{align}
\bPsi =
\left(
\begin{array}{c}
\bPsi_+ \\
\bPsi_-
\end{array}
\right),~~
\bPsi^\dagger =
\left(
\begin{array}{cc}
\bPsi_+^\dagger & \bPsi_-^\dagger
\end{array}
\right),
\end{align}
and
\begin{align}
&
\bPsi_\mu (\br) =
   \left(
     \begin{array}{c}
       \psi_{\mu \uparrow}(\br) \\
       \psi_{\mu \downarrow}(\br) \\
       \psi_{\mu \uparrow}^\dagger(\br) \\
       \psi_{\mu \downarrow}^\dagger(\br)
     \end{array}
   \right), \cr
%~~
&
\bPsi_\mu^\dagger(\br) =
   \left(
     \begin{array}{cccc}
       \psi_{\mu \uparrow}^\dagger(\br) &  \psi_{\mu \downarrow}^\dagger(\br) & \psi_{\mu \uparrow}(\br)
       & \psi_{\mu \downarrow}(\br)
     \end{array}
   \right).
\end{align}
Their Heisenberg representations are written as
\begin{align}
\bPsi_\mu(\br,\tau) = \re^{\H\tau} \bPsi_\mu(\br) \re^{-\H\tau},~~
\bPsi_\mu^\dagger(\br,\tau) = \re^{\H\tau} \bPsi_\mu^\dagger(\br) \re^{-\H\tau}.
\label{eqn:Heisenberg}
\end{align}
In the absence of impurity scattering, the unperturbed Green's function is Fourier-transformed to
\begin{align}
\bG_0(\ri\om,\bk)
  = - \frac{ \ri\om + \epsilon_{\bsk}\brho_3 + \bDelta \brho_2 \bsigma_2 }
   { \omd + \epsilon_{\bsk}^2 + \Delta^2 },
\end{align}
where $\bsigma_\alpha$ is the Pauli matrix for the spin space and $\brho_\alpha$ is that for the
particle-hole space ($\alpha = 1,2$, and $3$ correspond to $x,y$, and $z$, respectively),
and $\epsilon_{\bsk} = \bk^2 / (2m_{\rm e}) - E_{\rm F}$.
The two-band superconductivity is expressed with another Pauli matrix $\btau_\alpha$ in
the band space as
\begin{align}
\bDelta = \frac{\Delta_+}{2} (1 + \btau_3) + \frac{\Delta_-}{2} (1 - \btau_3).
\end{align}
\par

We consider the magnetic (octupolar) scattering effect on $T_{\rm c}$ when the impurities are
distributed randomly in the two-band $s$-wave superconductor, which is expressed on
the right-hand side of the following linearized gap equation:
\cite{Fulde70,Koga10}
\begin{align}
\frac{8 T_{\rm c}}{\pi} \tau^S \log \frac{T_{\rm c}}{T_{{\rm c}0}}
\left(
\begin{array}{c}
\Delta_+ \\
\Delta_-
\end{array}
\right) =
\left(
\begin{array}{cc}
f_\omega & f_\Delta \\
f_\Delta & f_\omega
\end{array}
\right)
\left(
\begin{array}{c}
\Delta_+ \\
\Delta_-
\end{array}
\right),
\label{eqn:gap}
\end{align}
where $T_{\rm c}$ ($T_{{\rm c} 0}$) is the transition temperature in the presence (absence) of
impurities.
$f_\Delta$ and $f_\omega$ represent self-energies corresponding to the order parameter and
Matsubara frequency components, respectively.
On the basis of the unperturbed Hamiltonian $\H_{\rm C} + \H_{\rm I}$
in eqs.~(\ref{eqn:H-C}) and (\ref{eqn:H-I}),
the self-energy is obtained as
\begin{align}
\bSigma_S (\ri\om) = & - n_{\rm imp}T^2 \sum_{n \ne n'} \sum_{\omega_1\omega_2}
  \frac{1}{\ri\omega_1-\delta_n} \frac{1}{\ri\omega_2-\delta_{n'}} \cr
& \times J_S^2 \frac{1}{\Omega} \sum_\bsk
  \left( \bI_S \right)_{nn'} \bG_0 (\ri\om+\ri\omega_1-\ri\omega_2,\bk) \cr
&~~~~~~~~~~~~~~~~~~\times
  \left( \bI_S \right)_{n'n}
\end{align}
in the second Born approximation by $\H'$ in eq.~(\ref{eqn:H'}) for the $\Gamma_5$ magnetic
type of effective exchange scattering $\hat{I}_S$ in eq.~(\ref{eqn:I-S1}).
Here, $n_{\rm imp}$ is the impurity density, $T$ is the temperature ($k_{\rm B} = 1$), and
$\Omega$ represents the system volume.
By analogy with the optical phonon case, the inelastic impurity scattering leads to an effective
pairing interaction.
After calculating the self-energy terms,
\begin{align}
\Sigma_\Delta (\ri \om) = \frac{1}{8}{\rm Tr} \left[ \brho_2 \bsigma_2 \bSigma_S (\ri \om) \right],~~
\Sigma_\omega (\ri \om) = \frac{1}{8}{\rm Tr} \bSigma_S (\ri \om),
\end{align}
where $1/8$ is the normalization factor in the $\btau \otimes \brho \otimes \bsigma$ space
($\btau$ for the band space, $\brho$ for the particle-hole space and $\bsigma$ for the spin space),
we obtain
\begin{align}
& \pi T_{\rm c} \sum_l \frac{1}{| \om |} \Sigma_\Delta (\ri \om)
= - \frac{\pi}{8 T_{\rm c} \tau^S} f_\Delta (x) > 0, \cr
& \pi T_{\rm c} \sum_l \frac{\ri}{| \om |} \frac{\Sigma_\omega (\ri \om)}{\om}
= - \frac{\pi}{8 T_{\rm c} \tau^S} f_\omega (x) > 0,
\label{eqn:fxi}
\end{align}
where $x$ represents the energy difference between the $\Gamma_1$ singlet ground and
$\Gamma_5$ triplet excited states as
\begin{align}
x = \frac{\delta_{\Gamma_5} - \delta_{\Gamma_1}}{2T_{\rm c}}.
\end{align}
We define the lifetime due to the magnetic impurity scattering as
\begin{align}
\frac{1}{\tau^S} = 2 \pi n_{\rm imp} N_0 \left( \frac{J_S}{2} \right)^2,
\end{align}
where $N_0$ represents the density of electronic states at the Fermi energy.
In the absence of a magnetic field, both $f_\Delta (x)$ and $f_\omega (x)$ are independent of
$\theta$ and $\phi$ related to the local hybridization defined in eq.~(\ref{eqn:mix}).
In eq.~(\ref{eqn:gap}), the matrix gives the higher eigenvalue
$f(x) = - f_\Delta (x) + f_\omega (x) > 0$ for the higher \Tc~in the $s_\pm$-wave state
($\Delta_+ / \Delta_- = -1$).
The explicit representation of $f(x)$ is given in Appendix~B.
The $s_{++}$-wave state ($\Delta_+ / \Delta_- = 1$) is for the lower \Tc.
In the same manner, for the nonmagnetic (quadrupolar) scattering in Appendix~A, the gap
equation is given by
\begin{align}
\frac{8 T_{\rm c}}{\pi} \tau^Q \log \frac{T_{\rm c}}{T_{{\rm c}0}}
\left(
\begin{array}{c}
\Delta_+ \\
\Delta_-
\end{array}
\right) =
\left(
\begin{array}{cc}
f_\omega & - f_\Delta \\
- f_\Delta & f_\omega
\end{array}
\right)
\left(
\begin{array}{c}
\Delta_+ \\
\Delta_-
\end{array}
\right),
\end{align}
where $\tau^Q$ is the corresponding lifetime, which leads to the \Tc~increase in the
$s_{++}$-wave state (see Table~\ref{table:1}).
\par

%%%%%%%%%%%%%%%%%%%%%%%%%%%%%%%%%%%%%%%%%%%%%%%%%%%%%%%%%%%%%%%%%%%%%%%%%%%%%%%%%%%%%%%%%%%%%%%%%%%
\subsection{Magnetic field effect}
%%%%%%%%%%%%%%%%%%%%%%%%%%%%%%%%%%%%%%%%%%%%%%%%%%%%%%%%%%%%%%%%%%%%%%%%%%%%%%%%%%%%%%%%%%%%%%%%%%%
Next, we consider a magnetic field effect that is weak enough not to directly affect the 
superconducting order parameter and put aside the field coupling with conduction electrons.
In the present case, the Zeeman splitting of the impurity triplet states reduces the effective pairing
interaction and the $a_u$-$t_u$ scattering depends on the field direction.
Consequently, the reduced $T_{\rm c}$ exhibits the field orientation dependence.
\par

It is convenient to choose the quantization axis in the direction of the applied magnetic field $\bH$.
The Zeeman splitting is expressed as
\begin{align}
\langle \pm | - \bJ \cdot \bh | \pm \rangle = \mp h_{\rm t},~~
\langle 0 | - \bJ \cdot \bh | 0 \rangle = 0,
\label{eqn:triplet-h}
\end{align}
where $h_{\rm t} = (5/2) h$ and $h = |\bh|$
($\bh = g_J \mu_{\rm B} \bH$: $g_J$ is the Land{\'e} $g$ factor)
for the $\Gamma_5$ triplet, and $| n \rangle$ ($n = +,0,-$) is an eigenstate of
$- \bJ \cdot \bh$ expressed by a combination of the three $\Gamma_5$ states in
eq.~(\ref{eqn:Gamma5}).
For the interchange between the singlet and triplet states, the $T_\eta$ operators in
eq.~(\ref{eqn:I-S1}) correspondingly depend on the field directions
$(\bar{h}_x,\bar{h}_y,\bar{h}_z)$, as indicated in eqs.~(\ref{eqn:Szh}) and (\ref{eqn:S+h}).
The magnetic field effect only modifies
$f_{\xi = \omega, \Delta}$ in the gap equation~[eq.~(\ref{eqn:gap})] as
\begin{align}
f_\xi \rightarrow C_h f_\xi (x_0) + \frac{1}{2} ( 1 - C_h )[ f_\xi (x_+) + f_\xi (x_-) ],
\label{eqn:fxi-h}
\end{align}
where the energy difference $x_n$ ($n = +, 0, -$ for the excited triplet) and the anisotropy
coefficient $C_h$ are defined as 
\begin{align}
& x_n = \frac{\delta_n - \delta_{\Gamma_1}}{2 T_{\rm c}},
\label{eqn:x-n} \\
& C_h = {\rm Tr}_{\btau \bsigma}
\left[ \langle \Gamma_1 | \bI_S | 0 \rangle \langle 0 | \bI_S | \Gamma_1 \rangle \right],
\label{eqn:ch1}
\end{align}
respectively.
Here, ${\rm Tr}_{\btau \bsigma}$ indicates the trace of $[ \cdots ]$ in calculating with the
$4 \times 4$ matrices in eqs.~(\ref{eqn:sz}) and (\ref{eqn:s+}).

%%%%%%%%%%%%%%%%%%%%%%%%%%%%%%%%%%%%%%%%%%%%%%%%%%%%%%%%%%%%%%%%%%%%%%%%%%%%%%%%%%%%%%%%%%%%%%%%%%%
\section{Application}
\begin{figure}
\begin{center}
\includegraphics[width=7cm,clip]{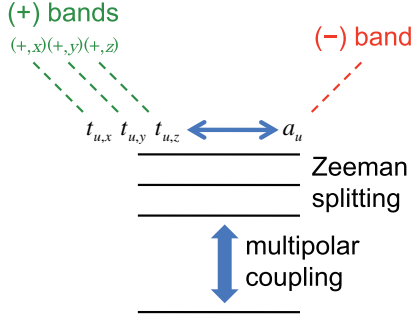}
\end{center}
\caption{Sketch of the inelastic multipolar exchange scattering in the $a_u$ and
threefold degenerate $t_u$ band case.
Each local $t_{u,\alpha}$ electron participates in the corresponding ($+,\alpha$) band
($\alpha = x,y,z$), while the local $a_u$ electron is transferred to the single ($-$) band.
}
\label{fig:2}
\end{figure}

%%%%%%%%%%%%%%%%%%%%%%%%%%%%%%%%%%%%%%%%%%%%%%%%%%%%%%%%%%%%%%%%%%%%%%%%%%%%%%%%%%%%%%%%%%%%%%%%%%%
In this section, we consider a case of the single $a_u$ and threefold degenerate $t_u$ bands
as an extension of the above two-band case (see Fig.~\ref{fig:2}).
The inclusion of the threefold degeneracy is required in the minimum model for the conservation
of the cubic symmetry.
We assume that these bands are combined with each other only through the impurity interband
scattering effect.
Since it is shown that \Tc~is not markedly field-orientation-dependent under the cubic symmetry,
we introduce a uniaxial anisotropy effect phenomenologically to check how \Tc~is affected by
the orbital (multipole) anisotropy of impurity scattering.
Finally, we give a comment on $H_{\rm c2} (T)$ in the vicinity of \Tc~that reflects the anisotropic
scattering.

%%%%%%%%%%%%%%%%%%%%%%%%%%%%%%%%%%%%%%
%%%%%%%%%%%%%%%%%%%%%%%%%%%%%%%%%%%%%%

%%%%%%%%%%%%%%%%%%%%%%%%%%%%%%%%%%%%%%%%%%%%%%%%%%%%%%%%%%%%%%%%%%%%%%%%%%%%%%%%%%%%%%%%%%%%%%%%%%%
\subsection{$a_u$ and threefold degenerate $t_u$ band case}
%%%%%%%%%%%%%%%%%%%%%%%%%%%%%%%%%%%%%%%%%%%%%%%%%%%%%%%%%%%%%%%%%%%%%%%%%%%%%%%%%%%%%%%%%%%%%%%%%%%
When a magnetic field is applied, it is generally observable that under the cubic symmetry,
$T_{\rm c}$ decreases as
\begin{align}
T_{\rm c} (h=0) - T_{\rm c} = a_2 h_{\rm t}^2 + a_4 h_{\rm t}^4,
\label{eqn:Tc-h}
\end{align}
where $h_{\rm t} = (5/2)h$ is the Zeeman splitting, $a_2$ is a positive constant, and
$a_4$ is the field-orientation-dependent parameter
\begin{align}
a_4 \propto {\rm const.} + (\bar{h}_x^4 + \bar{h}_y^4 + \bar{h}_z^4).
\end{align}
Here, we simplify the mixing of the local $t_u$ ($x,y,z$) orbitals and the $+$ band.
Instead, for the conservation of the cubic symmetry, we introduce the threefold degenerate
$t_u$-dominant bands denoted by ($+, \alpha$), where $\alpha = x, y, z$,  which hybridize with
the $f$-orbitals most strongly in the directions of three principal axes.
The corresponding $a_u$-$t_u$ scattering anisotropy in eq.~(\ref{eqn:ch2}) is given by
\begin{align}
& C_{h,x} \equiv C_h ( \theta = \frac{\pi}{2}, \phi = 0 )
= \frac{1}{6} + \frac{1}{2} \bar{h}_x^2, \cr
& C_{h,y} \equiv C_h ( \theta = \frac{\pi}{2}, \phi = \frac{\pi}{2} )
= \frac{1}{6} + \frac{1}{2} \bar{h}_y^2, \cr
& C_{h,z} \equiv C_h ( \theta = 0 )
= \frac{1}{6} + \frac{1}{2} \bar{h}_z^2.
\label{eqn:C-tu}
\end{align}
When the single $a_u$ and threefold degenerate $t_u$ bands are taken into account, the gap
equation [eq.~(\ref{eqn:gap})] is extended to the following eigenvalue problem:
\begin{align}
& \frac{ T_{\rm c} }{ T_{{\rm c} 0} } \log \frac{ T_{\rm c} }{ T_{{\rm c} 0} }
\left(
\begin{array}{c}
\Delta_{+,x} \\
\Delta_{+,y} \\
\Delta_{+,z} \\
\Delta_-
\end{array}
\right) = \bLambda
\left(
\begin{array}{c}
\Delta_{+,x} \\
\Delta_{+,y} \\
\Delta_{+,z} \\
\Delta_-
\end{array}
\right), \cr
&~~ \bLambda = \alpha_S
\left(
\begin{array}{cccc}
F_{\omega,x} & 0 & 0 & F_{\Delta,x} \\
0 & F_{\omega,y} & 0 & F_{\Delta,y} \\
0 & 0 & F_{\omega,z} & F_{\Delta,z} \\
F_{\Delta,x} & F_{\Delta,y} & F_{\Delta,z} & \ds{\sum_\alpha} F_{\omega,\alpha}
\end{array}
\right).
\label{eqn:Lambda}
\end{align}
Here, $\Delta_{+,\alpha}$ ($\alpha = x, y, z$) is the order parameter for the ($+,\alpha$)
band and no interband pairing is taken into account.
The magnetic scattering strength $\alpha_S$ is given by
\begin{align}
\alpha_S = \frac{\pi}{8 T_{{\rm c}0} \tau^S} \propto \frac{n_{\rm imp} N_0 J_S^2}{T_{{\rm c}0}}.
\end{align}
In the presence of a magnetic field, the matrix elements of $\bLambda$ for $\xi$
($= \omega, \Delta$) are given in the same manner as eq.~(\ref{eqn:fxi-h}):
\begin{align}
F_{\xi,\alpha} = C_{h,\alpha} f_\xi (x_0) + \frac{1}{2}(1 - C_{h,\alpha}) [ f_\xi (x_+) + f_\xi (x_-) ].
\label{eqn:Fxi-a}
\end{align}
The excited triplet energy levels are split as
\begin{align}
x_0 = \frac{\delta_0 - \delta_{\Gamma_1}}{2T_{\rm c}},~~
x_{\pm} = x_0 \mp x_h~~\left( x_h \equiv \frac{h_{\rm t}}{2 T_{\rm c}} \right),
\end{align}
owing to the Zeeman splitting $h_{\rm t} = (5/2)h$ of the $\Gamma_5$ triplet.
When the field direction is so rotated as to pass through the [111] axis, both $x$ and $y$
components are taken to be equivalent ($\bar{h}_x = \bar{h}_y$).
In Appendix~C, we show that the field-direction ($\bar{h}_z$)-dependent term is extracted
from $F_{\xi,\alpha}$. 
\par

\begin{figure}
\begin{center}
\includegraphics[width=7cm,clip]{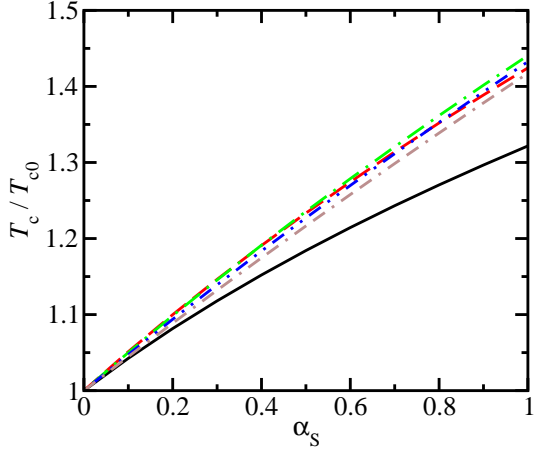}
\end{center}
\caption{$T_{\rm c} / T_{{\rm c}0}$ of the $s_\pm$-wave state as a function of the
interband magnetic scattering strength $\alpha_S$ for various values of the level splitting
$(\delta_{\Gamma_5} - \delta_{\Gamma_1}) / T_{{\rm c}0} =5$ (solid line), $10$ (dashed line),
$15$ (one-dashed and one-dotted line), $20$ (one-dashed and two-dotted line), and
$25$ (two-dashed and one-dotted line).
The same dependence is obtained for the interband nonmagnetic scattering in the $s_{++}$-wave
state.
}
\label{fig:3}
\end{figure}

The highest eigenvalue of $\bLambda$ determines $T_{\rm c}$.
When a magnetic field is absent ($x_h = 0$), $F_{\xi,\alpha} = f_{\xi,\alpha} = f_{\xi}$
in eq.~(\ref{eqn:f-xi-b}) leads to
\begin{align}
\frac{ T_{\rm c} }{ T_{{\rm c} 0} } \log \frac{ T_{\rm c} }{ T_{{\rm c} 0} } =
\alpha_S \left( 2 f_\omega + \sqrt{ f_\omega^2 + 3 f_\Delta^2 } \right).
\label{eqn:gap-f}
\end{align}
The $x_0$ dependence of this equation is similar to $f(x)$ in Fig.~\ref{fig:b1}
that corresponds to the two-band case.
The calculated $T_{\rm c}$ is plotted for various values of the singlet-triplet level splitting
$(\delta_{\Gamma_5} - \delta_{\Gamma_1}) / T_{{\rm c}0}$ in Fig.~\ref{fig:3}.
The $\alpha_S$ dependence is not sensibly affected by the large change in crystal field level for
$(\delta_{\Gamma_5} - \delta_{\Gamma_1}) \gtrsim  10T_{{\rm c}0}$.
The monotonic increase in $T_{\rm c}$ with $\alpha_S \propto n_{\rm imp}$ explains well
the Pr concentration $x$ dependence of $T_{\rm c}$ measured in
\LaPr~at the relatively small $x$ values where the Pr ions are
regarded as impurities.
\cite{Yogi06}
\par

%%%%%%%%%%%%%%%%%%%%%%%%%%%%%%%%%%%%%%
%%%%%%%%%%%%%%%%%%%%%%%%%%%%%%%%%%%%%%

In the presence of a magnetic field, the effect of $\eta_\xi$ in eq.~(\ref{eqn:eta}) appears at
the field-orientation-dependent \Tc~in eq.~(\ref{eqn:Tc-h}) such that
\begin{align}
a_4 \sim \frac{1}{2} - \bar{h}_z^2 + \frac{3}{2} \bar{h}_z^4,
\end{align}
where $\bar{h}_x = \bar{h}_y$ is retained.
This leads to
\begin{align}
\frac{a_{4[001]} - a_{4[111]}}{a_{4[110]} - a_{4[111]}} =4.
\end{align}
In the present model, we find that the difference ($T_{{\rm c}[001]} - T_{{\rm c}[111]}$)
is of the order of $h^4$, less than $10^{-4} T_{\rm c}$ even for a larger magnetic field $x_h > 1$ since $|a_4|$ is extremely smaller than $a_2$.
Therefore, $\eta_{\xi}$ is negligible and $T_{\rm c}$ can be regarded as isotropic, which is
determined by
\begin{align}
\frac{ T_{\rm c} }{ T_{{\rm c} 0} } \log \frac{ T_{\rm c} }{ T_{{\rm c} 0} } =
\alpha_S \left( 2 F_\omega + \sqrt{ F_\omega^2 + 3 F_\Delta^2 } \right)
\end{align}
with the same form as eq.~(\ref{eqn:gap-f}).
The decrease in $T_{\rm c}$ with increasing $h$ comes from
\begin{align}
F(x_0,x_h) - f(x_0) = \frac{1}{3} f''(x_0) x_h^2 < 0,
\end{align}
where $F = - F_\Delta + F_\omega$ and $f = - f_\Delta + f_\omega > 0$,
calculated using eq.~(\ref{eqn:Fx0}).
It should be noted that \Tc~could be increased by applying a magnetic field
if $f(x)$ had some features to satisfy $f''(x_0) > 0$.

%%%%%%%%%%%%%%%%%%%%%%%%%%%%%%%%%%%%%%%%%%%%%%%%%%%%%%%%%%%%%%%%%%%%%%%%%%%%%%%%%%%%%%%%%%%%%%%%%%%
\subsection{Uniaxial anisotropy effect}
%%%%%%%%%%%%%%%%%%%%%%%%%%%%%%%%%%%%%%%%%%%%%%%%%%%%%%%%%%%%%%%%%%%%%%%%%%%%%%%%%%%%%%%%%%%%%%%%%%%
As discussed above, no distinct feature is found in the field orientation dependence of $T_{\rm c}$
in the rotation of the field direction.
This is due to the equivalency of $x$, $y$, and $z$ components in the orbital symmetry and the
constant $a_2$ in eq.~(\ref{eqn:Tc-h}).
Here, we check how the $T_{\rm c}$ deviation is increased by lowering the crystal field symmetry
from $O_h$.
To show this explicitly, we introduce a uniaxially anisotropic deviation from $O_h$.
Consequently, $a_2$ has a linear dependence on $\bar{h}_z^2$,
which leads to the field angle dependence as $\cos 2 \theta_h$ ($\cos \theta_h = \bar{h}_z$),
reflecting the twofold symmetry.
\par

\begin{figure}
\begin{center}
\includegraphics[width=7cm,clip]{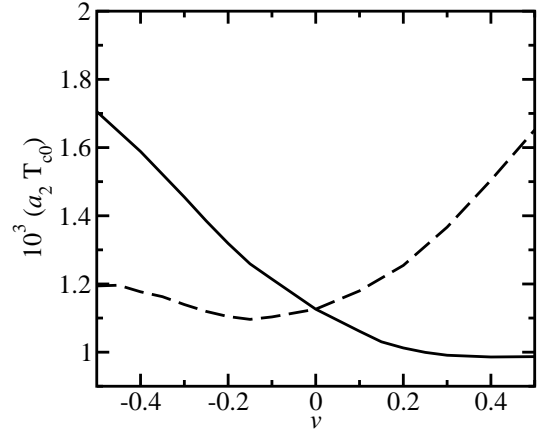}
\end{center}
\caption{Anisotropy $v$ dependence of $a_2 (\theta_h)$ for $\alpha_S = 1$ and
$(\delta_0 - \delta_{\Gamma_1}) / T_{{\rm c}0} = 10$.
The solid and dashed lines are the plots for $\bh \| [001]$ ($\theta_h = 0$) and
$\bh \| [110]$ ($\theta_h = \pi / 2$), respectively.
$a_2$ exhibits the $\cos 2 \theta_h$ oscillation between $a_2 (0)$ and $a_2 (\pi / 2)$.
}
\label{fig:4}
\end{figure}

Here, we represent the uniaxial anisotropy by the single phenomenological parameter $v$ in the
gap equation [eq.~(\ref{eqn:Lambda})], modifying $F_{\xi,\alpha}$ as
\begin{align}
F_{\xi,\alpha} \rightarrow (1- v) F_{\xi,\alpha}~~(\alpha = x,y),~~
F_{\xi,z} \rightarrow (1 + 2v ) F_{\xi,z}.
\end{align}
For the impurity states, only the Zeeman splitting $x_h$ is considered.
We assume that the main contribution to $v$ comes from the change in local orbital hybridization
amplitude owing to the lowering of the symmetry.
If the limits $v \rightarrow 1$ and $\alpha_S \rightarrow \alpha_S / 3$ are considered,
eq.~(\ref{eqn:Lambda}) is reduced to the two-band case in eq.~(\ref{eqn:gap}).
Around the $z$-axis, we find that $a_2$ in eq.~(\ref{eqn:Tc-h}) is independent of
$(\bar{h}_x, \bar{h}_y)$ since the fourfold symmetry is conserved.
In the following argument, we calculate 
\begin{align}
a_2(\theta_h) = \frac{1}{2} \left( a_{2[001]} + a_{2[110]} \right)
+ \frac{1}{2} \left( a_{2[001]} - a_{2[110]} \right) \cos 2 \theta_h
\label{eqn:Tcangle}
\end{align}
for various $v$ values, keeping $\bar{h}_x = \bar{h}_y$ in eq.~(\ref{eqn:Lambda}).
As it is expected, the $T_{\rm c}$ deviation becomes more distinct as the crystal field anisotropy
$v$ increases.
Figure~\ref{fig:4} shows how $a_2$ increases with $|v|$ for both
$\bh \| [001]$ ($\theta_h = 0$) and $\bh \| [110]$ ($\theta_h = \pi / 2$).
The maximum of $T_{\rm c}$ (minimum of $a_2$) is given by $T_{{\rm c}[001]}$ when $v > 0$
and by $T_{{\rm c}[110]}$ when $v < 0$.
It is necessary for the larger amplitude of oscillation in eq.~(\ref{eqn:Tcangle}) to introduce
a stronger anisotropy.
For $v=0.5$, the difference $(a_{2[110]} - a_{2[001]})$ is estimated as
$\sim 10^{-3} / T_{\rm c}$ in Fig.~\ref{fig:4}.
This indicates that the amplitude of $T_{\rm c}$ oscillation ($\propto h^2$) is about
$0.1\%$ of $T_{\rm c}$ at $h_{\rm t} / T_{\rm c} \simeq 1$ and $1\%$ at
$h_{\rm t} / T_{\rm c} \simeq 3$ ($h_{\rm t}$ represents the Zeeman splitting of the excited triplet).
If the crystal field anisotropy is much smaller, it is more difficult to obtain the field angle $\theta_h$
dependence of $T_{\rm c}$.
We note that $\eta_\xi$ in eq.~(\ref{eqn:eta}) is responsible for the $\theta_h$ dependence,
so that $T_{\rm c}$ at $\bar{h}_z = 1 / \sqrt{3}$ for $v \ne 0$ ($T_{{\rm c}[111]}$ in the cubic
symmetry for $v =0$) directly reflects the crystal field anisotropy ($t_u$-orbital anisotropy) since
$\eta_\xi$ vanishes at $\bar{h}_z = 1 / \sqrt{3}$.
Therefore, the anisotropy of multipolar scattering  itself appears in the $T_{\rm c}$ deviation from
$T_{\rm c}(\bar{h}_z = 1 / \sqrt{3})$.
\par

%%%%%%%%%%%%%%%%%%%%%%%%%%%%%%%%%%%%%%
%%%%%%%%%%%%%%%%%%%%%%%%%%%%%%%%%%%%%%

In a real system, this anisotropy effect can be observed as the field orientation dependence
of the upper critical field $h_{{\rm c}2}$ near $T_{\rm c}$ at $h=0$.
We can estimate $h_{{\rm c}2}$ in the framework of the conventional Ginzburg-Landau (GL)
theory, taking into account a diamagnetic effect in the GL expansion of a free energy.
\cite{Abrikosov57}
$h_{{\rm c}2}$ is given by $\Delta t $ ($= 1 - T / T_{\rm c}$) expansion as
\cite{Ichioka97}
\begin{align}
h_{{\rm c}2} = c_1 \Delta t + c_2 (\Delta t)^2 + \cdots,
\label{eqn:hc2}
\end{align}
with $c_1 >0$.
In the present case, the effective pairing interaction depends on the crystal field energy levels
coupled to the magnetic field, and $T_{\rm c}$ is reduced as
\begin{align}
T_{\rm c} = T_{\rm c}(0) - a_2 (\theta_h) h_{{\rm c}2}^2~~
\left[ T_{\rm c}(0) \equiv T_{\rm c}(h=0) \right].
\end{align}
Substituting it for $T_{\rm c}$ in eq.~(\ref{eqn:hc2}), we obtain
\begin{align}
h_{{\rm c}2} \simeq c_1 \left[ 1 - \frac{T}{T_{\rm c}(0)} \right]
+ \left[ c_2 - \frac{c_1^3}{T_{\rm c}(0)} a_2 (\theta_h) \right] \left[ 1 - \frac{T}{T_{\rm c}(0)} \right]^2.
\end{align}
When the cubic symmetry is conserved, $h_{{\rm c}2}$ is invariant against the rotation of the field
direction since $a_2 (\theta_h)$ is a constant for $v=0$.
If a uniaxial anisotropy is applied, for instance, by pressure measurement, the
field-angle-dependent $h_{{\rm c}2} (T)$ lines could be observed in the superconducting phase.
This is more promising for a large $c_1$ that is related to the GL parameter.
This provides us with conclusive evidence of the multiband picture proposed in our theory.
The multipole moments in the $f$-electron states play an important role in the anisotropic
$h_{{\rm c}2}(T)$, which is analogous to the quadrupole ordering transition temperature
$T_{\rm Q} (H)$ with the field orientation dependence, as observed in PrPb$_3$
\cite{Tayama01,Onimaru04}
or as predicted for CeB$_6$ in a high magnetic field.
\cite{Shiina02}

%%%%%%%%%%%%%%%%%%%%%%%%%%%%%%%%%%%%%%%%%%%%%%%%%%%%%%%%%%%%%%%%%%%%%%%%%%%%%%%%%%%%%%%%%%%%%%%%%%%
\section{Conclusion}
%%%%%%%%%%%%%%%%%%%%%%%%%%%%%%%%%%%%%%%%%%%%%%%%%%%%%%%%%%%%%%%%%%%%%%%%%%%%%%%%%%%%%%%%%%%%%%%%%%%
We have studied a magnetic field effect on \Tc~increased by inelastic scattering for
the singlet-triplet configuration in impurities.
The \Tc~increase can be expected for either the interband magnetic scattering in
the $s_\pm$-wave state or the nonmagnetic scattering in the $s_{++}$-wave state.
We have focused on the $\Gamma_5$-type octupolar exchange scattering in the former.
The same argument is applicable to the $\Gamma_5$-type quadrupolar exchange scattering
in the latter.
Owing to the anisotropy of multipolar scattering [$\eta_\xi$ in eq.~(\ref{eqn:eta})],
\Tc~exhibits the field orientation dependence.
Since the \Tc~deviation ($\propto h^4$) is very small in the cubic symmetric environment,
we have introduced a uniaxial anisotropy into the degenerate $t_u$ orbitals to clarify the
multipolar scattering effect on \Tc.
The multipolar scattering anisotropy itself appears in the twofold symmetric oscillation of \Tc~with
the rotation of the field direction, and the amplitude of oscillation increases proportionally to
$h^2$.
This can be confirmed by observing the splitting of an $H_{{\rm c}2}(T)$ line near \Tc~with a
change in field angle, although a uniaxial anisotropic field is required to produce a clear
difference between the two field directions, as discussed in \S4.2.
\par

Thus, we have clarified the roles of multipole degrees of freedom in multiband superconductors.
The key is that the local orbital exchange scattering is directly connected to the interband
scattering via the hybridization of $f$-orbitals with each band.
The local orbital anisotropy in eq.~(\ref{eqn:eta}) appears clearly at the
field-orientation-dependent \Tc, which is related to the field orientation dependence in
eq.~(\ref{eqn:ch2}) for the octupolar scattering.
This is not expected for the spin exchange scattering in a single band.
If each $\bt_{\eta}$ ($\eta = z, \pm$) is replaced by the Pauli matrix $\bsigma_\eta$
($\bsigma_\pm = \bsigma_x \pm \ri \bsigma_y$) in eq.~(\ref{eqn:I-S2}), the calculated $C_h$
becomes a constant, indicating that the spin exchange scattering is invariant against the rotation
of the field direction.
Therefore, the field-orientation-dependent \Tc, i.e., the anisotropic $H_{{\rm c}2}$, would be
positive evidence for the multiband picture proposed here.
\par

In a real system, the quadrupolar scattering coexists with the octupolar scattering for the
$O_h$ $\Gamma_1$ singlet and $\Gamma_5$ triplet configurations.
\cite{Koga10}
The quadrupolar scattering strength $\alpha_Q$ is defined as $\alpha_S$.
Unlike the magnetic case, the interband nonmagnetic scattering suppresses \Tc~in
the $s_\pm$-wave state, which is due to the sign reversal of $F_{\Delta,\alpha}$ 
($\alpha = x,y,z$) in the gap equation [eq.~(\ref{eqn:Lambda})].
The competition of magnetic and nonmagnetic scattering effects can be taken into account by
modifying
\begin{align}
\alpha_S F_{\omega, \alpha} \rightarrow (\alpha_S + \alpha_Q ) F_{\omega,\alpha},~~
\alpha_S F_{\Delta, \alpha} \rightarrow (\alpha_S - \alpha_Q ) F_{\Delta,\alpha}
\end{align}
in eq.~(\ref{eqn:Lambda}).
The calculated \Tc~decreases linearly with $\alpha_Q / \alpha_S$ in the $s_\pm$-wave state
($\Delta_{+,\alpha} \Delta_- < 0$).
For the strong spin-orbit coupling, $\alpha_Q / \alpha_S = 1/9$ is derived from the Anderson
model including the effective exchange scattering of the $J = 5/2$
($\Gamma_7 \oplus \Gamma_8$ for $O_h$ in Appendix~A) electrons due to a single impurity.
\cite{Koga10,Koga06}
This ratio of scattering strengths leads to only an approximately $10\%$ reduction in \Tc.
Therefore, the \Tc~increase holds for the dominant magnetic scattering in the
$s_\pm$-wave state.
On the other hand, in the $s_{++}$-wave state ($\Delta_{+,\alpha} \Delta_- > 0$),
the nonmagnetic scattering contributes to the \Tc~increase for $\alpha_Q \gg \alpha_S$.
\cite{Koga10}
Experimentally, it has not been established which scattering is more dominant, magnetic or 
nonmagnetic, as the Pr impurity effect on the LaOs$_4$Sb$_{12}$ superconductor.
We would like to point out that \Tc~can also be suppressed by intraband magnetic scattering
that corresponds to the spin exchange scattering in the single $a_u$ band that we have
neglected.
Since the $a_u$ electrons are coupled only to the excited triplet in the strong spin-orbit coupling
case,
\cite{Koga06}
the intraband scattering effect on \Tc~is negligibly small.
\par

As mentioned in \S1, the $T_h$ symmetry is another feature of the skutterudites, leading to
the mixing of the $O_h$ $\Gamma_4$ and $\Gamma_5$ wave functions of the triplet states as
\cite{Shiina04a,Shiina04b}
\begin{align}
|\Gamma_4^{(2)} \rangle = \sqrt{1 - d^2} |\Gamma_5 \rangle + d |\Gamma_4 \rangle,
\end{align}
where $d$ represents the deviation from the $O_h$ symmetry.
Since $d$ is relatively small, the $T_h$ effect modifies $\eta_\xi$ slightly in
eq.~(\ref{eqn:eta}).
On the other hand, the magnetic field couples the ground state and one of the triplet states
via the Van Vleck process, which shifts the energy difference $x_n$ in eq.~(\ref{eqn:x-n})
by $\sim d^2 h^2 / (\delta_0 - \delta_{\Gamma_1})$.
Thus, the $T_h$ deviations from $O_h$ only give minor corrections to the present results
as long as the magnetic field is not large.
It should be noted that the magnetic field effect on \Tc~is more sensitive to the local hybridization
of $f$-electrons with conduction bands that we have assumed to be the strongest in the
directions of three principal axes, leading to the maximum or minimum \Tc~in the field orientation
dependence.
\par

Finally, we would like to refer to a few experimental studies to elucidate the crucial roles of the
localized Pr $4f$-electrons in the superconductivity.
In Pr$_x$Os$_4$Sb$_{12}$ synthesized under a high pressure, the resistivity and magnetization
data show the close correlation between the Pr singlet-triplet energy splitting and \Tc.
\cite{Tanaka09}
The recent nuclear magnetic resonance study indicates the relevance of magnetic multipole
(dipole and octupole) fluctuations for mass enhancement in \Pr.
\cite{Tou10}
For comparison with our scenario in the future, systematic experimental studies of
$H_{{\rm c}2}(T)$ in \LaPr~are highly desired, following the detailed analysis of $H_{{\rm c}2}(T)$
in \Pr~reported previously.
\cite{Measson04}
The observation of the anisotropic $H_{{\rm c}2}(T)$ is worth testing under the uniaxial pressure
that could enhance the Pr multipolar scattering anisotropy in a magnetic field.

%%%%%%%%%%%%%%%%%%%%%%%%%%%%%%%%%%%%%%%%%%%%%%%%%%%%%%%%%%%%%%%%%%%%%%%%%%%%%%%%%%%%%%%%%%%%%%%%%%%
\acknowledgement
%%%%%%%%%%%%%%%%%%%%%%%%%%%%%%%%%%%%%%%%%%%%%%%%%%%%%%%%%%%%%%%%%%%%%%%%%%%%%%%%%%%%%%%%%%%%%%%%%%%
This work is supported by a Grant-in-Aid for Scientific Research (No. 20540353) from the Japan
Society for the Promotion of Science.
One of the authors (H. K.) is supported by a Grant-in-Aid for Scientific Research on Innovative
Areas ``Heavy Electrons'' (No. 20102008)
from the Ministry of Education, Culture, Sports, Science, and Technology, Japan.

%%%%%%%%%%%%%%%%%%%%%%%%%%%%%%%%%%%%%%%%%%%%%%%%%%%%%%%%%%%%%%%%%%%%%%%%%%%%%%%%%%%%%%%%%%%%%%%%%%%
\appendix
\section{Interband Impurity Scattering with $a_u$-$t_u$ Orbital Exchange}
%%%%%%%%%%%%%%%%%%%%%%%%%%%%%%%%%%%%%%%%%%%%%%%%%%%%%%%%%%%%%%%%%%%%%%%%%%%%%%%%%%%%%%%%%%%%%%%%%%%
The single $f$-electron states with the $J = 5/2$ ($J_z = 5/2, 3/2, \cdots, -5/2$) total angular
momentum are classified into the $O_h$ symmetric states as
\cite{Onodera66}
\begin{align}
& \left\{
\begin{array}{l}
| \Gamma_{8,3/2} \rangle = - \sqrt{\ds{\frac{1}{6}}} | 3/2 \rangle - \sqrt{\ds{\frac{5}{6}}} | -5/2 \rangle, \\
| \Gamma_{8,1/2} \rangle = | 1/2 \rangle, \\
| \Gamma_{8,-1/2} \rangle = - | -1/2 \rangle, \\
| \Gamma_{8,-3/2} \rangle = \sqrt{\ds{\frac{1}{6}}} | -3/2 \rangle + \sqrt{\ds{\frac{5}{6}}} | 5/2 \rangle,
\end{array}
\right. \\
& \left\{
\begin{array}{l}
| \Gamma_{7,1/2} \rangle = \sqrt{\ds{\frac{5}{6}}} | -3/2 \rangle - \sqrt{\ds{\frac{1}{6}}} | 5/2 \rangle, \\
| \Gamma_{7,-1/2} \rangle = \sqrt{\ds{\frac{5}{6}}} | 3/2 \rangle - \sqrt{\ds{\frac{1}{6}}} | -5/2 \rangle. \\
\end{array}
\right.
\end{align}
The fourfold degenerate $\Gamma_8$ wave functions mix with the threefold degenerate $t_u$
($x,y,z$) orbitals as
\begin{align}
& | \Gamma_{8,-3/2} \rangle \leftrightarrow \frac{1}{\sqrt{2}}
( | x, \downarrow \rangle - \ri | y, \downarrow \rangle ), \cr
& | \Gamma_{8,3/2} \rangle \leftrightarrow - \frac{1}{\sqrt{2}}
( | x, \uparrow \rangle + \ri | y, \uparrow \rangle ), \cr
& | \Gamma_{8,1/2} \rangle \leftrightarrow \frac{1}{\sqrt{3}}
   \left[ \sqrt{2} | z, \uparrow \rangle - \frac{1}{\sqrt{2}}
   ( | x, \downarrow \rangle + \ri | y, \downarrow \rangle ) \right], \cr
& | \Gamma_{8,-1/2} \rangle \leftrightarrow \frac{1}{\sqrt{3}}
   \left[ \sqrt{2} | z, \downarrow \rangle + \frac{1}{\sqrt{2}}
   ( | x, \uparrow \rangle - \ri | y, \uparrow \rangle ) \right]. \cr
&
\end{align}
In the same manner, the doubly degenerate $\Gamma_7$ electrons are directly transferred to
the single $a_u$ ($xyz$) orbital as
\begin{align}
& | \Gamma_{7,1/2} \rangle \leftrightarrow \ri~| xyz, \uparrow \rangle, \cr
& | \Gamma_{7,-1/2} \rangle \leftrightarrow \ri~| xyz, \downarrow \rangle.
\end{align}
In eq.~(\ref{eqn:I-S1}), the $T_\eta$ octupole operators couple both the $\Gamma_1$ and
$\Gamma_5$ states, and their matrix expressions are given by
\cite{Koga10}
\begin{align}
& T_z =
\left(
\begin{array}{cccc}
0 & 0 & 1 & 0 \\
0 & 0 & 0 & 0 \\
1 & 0 & 0 & 0 \\
0 & 0 & 0 & 0
\end{array}
\right),
\label{eqn:Sz} \\
& T_+ = T_-^\dagger = \sqrt{2}
\left(
\begin{array}{cccc}
0 & 0 & 0 & 1 \\
- 1 & 0 & 0 & 0 \\
0 & 0 & 0 & 0 \\
0 & 0 & 0 & 0
\end{array}
\right).
\label{eqn:S+}
\end{align}
The interchange of the singlet ground and triplet excited states occurs via exchange in
$\Gamma_7$ and $\Gamma_8$ local electrons hybridizing with the $a_u$ and $t_u$ bands,
respectively.
Introducing
$\bpsi = (\psi_{+ \uparrow}~\psi_{+ \downarrow}~\psi_{- \uparrow}~\psi_{- \downarrow})^t$
for the electrons, where $\psi_{\mu \sigma}$ is the field operator for the $\mu = +$ ($t_u$) and
$\mu = -$ ($a_u$) bands with the spin $\sigma$ ($=\uparrow, \downarrow$),
we present the typical octupolar exchange operators with the following matrix expressions:
\cite{Koga10}
\begin{align}
& \bt_z = \ri \frac{1}{2}
\left(
\begin{array}{cccc}
0 & 0 & - c & - s_2 e^{- \ri \phi} \\
0 & 0 & s_2 e^{\ri \phi} & - c \\
c & - s_2 e^{- \ri \phi} & 0 & 0 \\
s_2 e^{\ri \phi} & c & 0 & 0
\end{array}
\right),
\label{eqn:sz} \\
& \bt_+ = \bt_-^\dagger = \ri \frac{1}{2}
\left(
\begin{array}{cccc}
0 & 0 & - \sqrt{3} s_1 e^{\ri \phi} & c \\
0 & 0 & 0 & - s_2 e^{\ri \phi} \\
s_2 e^{\ri \phi} & c & 0 & 0 \\
0 & \sqrt{3} s_1 e^{\ri \phi} & 0 & 0
\end{array}
\right).
\label{eqn:s+}
\end{align}
In the matrix elements, $c$, $s_1$, and $s_2$ are given by
\begin{align}
c = \sqrt{\frac{2}{3}} \cos \theta,~~s_1 = \frac{1}{\sqrt{2}} \sin \theta,~~
s_2 = \frac{1}{\sqrt{6}} \sin \theta,
\end{align}
where $\theta$ takes arbitrary values as well as $\phi$, which are related to the mixing of the
local $t_u$($x,y,z$) symmetric orbitals and the $+$ band shown as
\begin{align}
\langle x | + \rangle : \langle y | + \rangle : \langle z | + \rangle
= \sin \theta \cos \phi : \sin \theta \sin \phi : \cos \theta.
\label{eqn:mix}
\end{align}
Here, $| + \rangle$ represents a partial wave of the $+$ band electrons with the $t_u$
symmetry at an impurity site.
The details of impurity scattering due to such a multipole as the octupole are described for the
$f^2$ singlet-triplet configuration in the previous paper.
\cite{Koga10}
\par

In the presence of a magnetic field, the $T_\eta$ operators are expressed as
\begin{align}
& T_z  =
\left(
\begin{array}{cccc}
0 & b & (a_+ - a_-) & -b \\
b & 0 & 0 & 0 \\
(a_+ - a_-) & 0 & 0 & 0 \\
-b & 0 & 0 & 0
\end{array}
\right),
\label{eqn:Szh} \\
& T_+ =
\left(
\begin{array}{cccc}
0 & \sqrt{2} a_- e^{\ri \phi_h} & \sqrt{2} b e^{\ri \phi_h} & \sqrt{2} a_+ e^{\ri \phi_h} \\
- \sqrt{2} a_+ e^{\ri \phi_h} & 0 & 0 & 0 \\
\sqrt{2} b e^{\ri \phi_h} & 0 & 0 & 0 \\
- \sqrt{2} a_- e^{\ri \phi_h} & 0 & 0 & 0
\end{array}
\right),
\label{eqn:S+h}
\end{align}
on the basis of ($\Gamma_1, +, 0, -$), where ($+,0,-$) denote the excited triplet states that satisfy
eq.~(\ref{eqn:triplet-h}).
The parameters in each matrix element are magnetic-field-dependent and are defined as
\begin{align}
& a_\pm = \frac{1}{2} ( 1 \pm \bar{h}_z ),~~b = \frac{1}{\sqrt{2}} \sqrt{\bar{h}_x^2 + \bar{h}_y^2},~~
\tan \phi_h = \frac{\bar{h}_y}{\bar{h}_x},
\end{align}
where $\bh / h \equiv (\bar{h}_x, \bar{h}_y, \bar{h}_z)$.
Equations~(\ref{eqn:Szh}) and (\ref{eqn:S+h}) for $\bar{h}_z = 1$ correspond to
eqs.~(\ref{eqn:Sz}) and (\ref{eqn:S+}), respectively.
Using the matrices in eqs.~(\ref{eqn:sz}) and (\ref{eqn:s+}) to calculate
\begin{align}
\langle \Gamma_1 | \bI_S | 0 \rangle =
(a_+ - a_-) \bt_z + \frac{1}{\sqrt{2}} b e^{\ri \phi_h} \bt_-
+ \frac{1}{\sqrt{2}} b e^{- \ri \phi_h} \bt_+
\label{eqn:I-S2}
\end{align}
in eq.~(\ref{eqn:ch1}), we obtain
\begin{align}
C_h =& \left( \frac{1}{6} + a_+ a_- \right) \sin^2 \theta
+ \left( \frac{2}{3} - 2 a_+ a_- \right) \cos^2 \theta \cr
& + 2 \sqrt{a_+ a_-} (a_+ - a_-) \sin \theta \cos \theta \cos (\phi_h - \phi) \cr
&
+ (a_+ a_-) \sin^2 \theta \cos 2(\phi_h - \phi),
\label{eqn:ch2}
\end{align}
which depends on the details of the hybridization of local $f$-electron states with the
$+$ band represented by $\theta$ and $\phi$ in eq.~(\ref{eqn:mix}).
We note the following equality:
\begin{align}
\sum_{n=0,\pm} {\rm Tr}_{\btau \bsigma}
\left[ \langle \Gamma_1 | \bI_S | n \rangle \langle n | \bI_S | \Gamma_1 \rangle \right] = 1.
\end{align}
\par

To compare the magnetic and nonmagnetic exchange properties, we also present
the $\Gamma_5$-type quadrupole operators
\begin{align}
& Q_z =
\left(
\begin{array}{cccc}
0 & 0 & - \ri & 0 \\
0 & 0 & 0 & 0 \\
\ri & 0 & 0 & 0 \\
0 & 0 & 0 & 0
\end{array}
\right), \\
& Q_+ = Q_-^\dagger =  - \sqrt{2}
\left(
\begin{array}{cccc}
0 & 0 & 0 & 1 \\
1 & 0 & 0 & 0 \\
0 & 0 & 0 & 0 \\
0 & 0 & 0 & 0
\end{array}
\right),
\end{align}
which correspond to the $T_\eta$ octupole operators in eqs.~(\ref{eqn:Sz}) and (\ref{eqn:S+}),
respectively, and the quadrupolar exchange operators
\begin{align}
& \bq_z = - \frac{1}{2}
\left(
\begin{array}{cccc}
0 & 0 & c & s_2 e^{- \ri \phi} \\
0 & 0 & - s_2 e^{\ri \phi} & c \\
c & - s_2 e^{- \ri \phi} & 0 & 0 \\
s_2 e^{\ri \phi} & c & 0 & 0
\end{array}
\right), \\
& \bq_+ = \bq_-^\dagger = \ri \frac{1}{2}
\left(
\begin{array}{cccc}
0 & 0 & \sqrt{3} s_1 e^{\ri \phi} & -c \\
0 & 0 & 0 & s_2 e^{\ri \phi} \\
s_2 e^{\ri \phi} & c & 0 & 0 \\
0 & \sqrt{3} s_1 e^{\ri \phi} & 0 & 0
\end{array}
\right),
\end{align}
which correspond to the octupolar exchange operators ($\bt_\eta$) in eqs.~(\ref{eqn:sz}) and
(\ref{eqn:s+}), respectively.
\cite{Koga10}
Let us take $s_1 = s_2 = 0$ to simplify $\bt_z$ and $\bq_z$ as
\begin{align}
&
\bt_z = \frac{c}{2}
\left(
\begin{array}{cccc}
0 & 0 & - \ri & 0 \\
0 & 0 & 0 & - \ri \\
\ri & 0 & 0 & 0 \\
0 & \ri & 0 & 0
\end{array}
\right), \cr
%~~
&
\bq_z = - \frac{c}{2}
\left(
\begin{array}{cccc}
0 & 0 & 1 & 0 \\
0 & 0 & 0 & 1 \\
1 & 0 & 0 & 0 \\
0 & 1 & 0 & 0
\end{array}
\right).
\end{align}
They express the spin-independent scattering with only orbital exchange.
In the band space, it is clear that the former $\btau_2$ type is magnetic and the latter $\btau_1$
type is nonmagnetic.
\cite{Koga10}
It is easy to check whether the effective paring interaction mediated by such impurity scattering
is attractive or repulsive in the $s_\pm$-wave state as follows.
Since the $s_\pm$-wave state is expressed by $\Delta \btau_3 \brho_2 \bsigma_2$ in the
$\btau \otimes \brho \otimes \bsigma$ space, the $\btau_2$-type magnetic scattering satisfies
\begin{align}
\btau_2 (\Delta \btau_3 \brho_2 \bsigma_2) \btau_2 = - \Delta \btau_3 \brho_2 \bsigma_2,
\end{align}
which is used for deriving $f_\Delta$ in eq.~(\ref{eqn:gap}).
The sign reversal of $\Delta$ indicates that the impurity-mediated pairing interaction is attractive.
On the contrary, it becomes repulsive for the $\btau_1$-type nonmagnetic scattering since
\begin{align}
\btau_1 \brho_3 (\Delta \btau_3 \brho_2 \bsigma_2) \btau_1 \brho_3
= \Delta \btau_3 \brho_2 \bsigma_2
\end{align}
results in the absence of sign reversal of $\Delta$ owing to $\brho_3$ in the particle-hole space
that accompanies the nonmagnetic scattering.
In a similar analysis, one can confirm that the nonmagnetic (magnetic) scattering leads to the
attractive (repulsive) interaction effectively for the $s_{++}$-wave pairing.
Therefore, \Tc~can be increased by either the magnetic scattering in the
$s_\pm$-wave state or the nonmagnetic scattering in the $s_{++}$-wave state in the case of
interband impurity scattering with the $a_u$-$t_u$ orbital exchange.

%%%%%%%%%%%%%%%%%%%%%%%%%%%%%%%%%%%%%%%%%%%%%%%%%%%%%%%%%%%%%%%%%%%%%%%%%%%%%%%%%%%%%%%%%%%%%%%%%%%
\section{Function as the Impurity Effect on $T_{\rm c}$}
%%%%%%%%%%%%%%%%%%%%%%%%%%%%%%%%%%%%%%%%%%%%%%%%%%%%%%%%%%%%%%%%%%%%%%%%%%%%%%%%%%%%%%%%%%%%%%%%%%%

%%%%%%%%%%%%%%%%%%%%%%%%%%%%%%%%%%%%%%
\begin{figure}
\begin{center}
\includegraphics[width=7cm,clip]{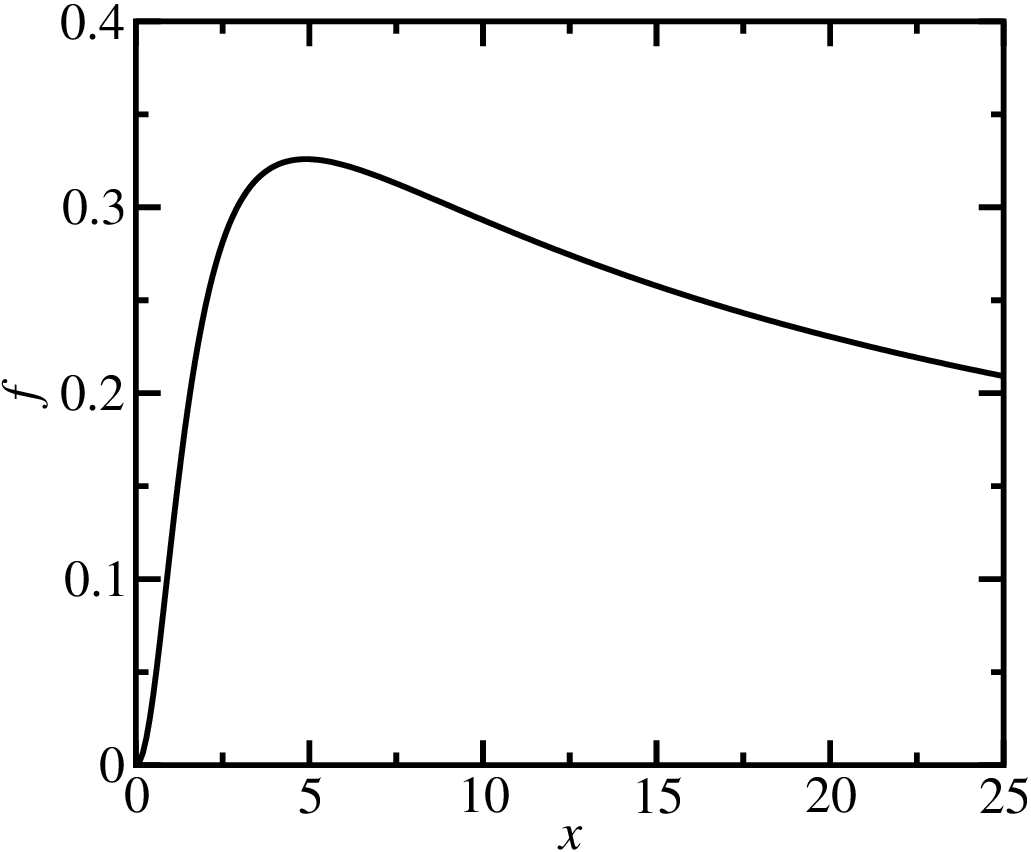}
\end{center}
\caption{Plot of $f(x)$.
}
\label{fig:b1}
\end{figure}
%%%%%%%%%%%%%%%%%%%%%%%%%%%%%%%%%%%%%%

Each gap equation presented in this paper gives the level-splitting $x$-dependent $T_{\rm c}$
determined by $f(x) = - f_\Delta (x) + f_\omega (x)$ that is defined in eq.~(\ref{eqn:fxi}).
For the calculation of $f(x)$ plotted in Fig.~\ref{fig:b1}, the necessary formulas are arranged here:
\cite{Fulde70,Koga10}
\begin{align}
& f_\Delta (x) = - \frac{\tanh x}{x} + A(x) - \frac{1}{2} B(x), \cr
& f_\omega (x) = - 1 + \tanh^2{x} - \frac{1}{2} B(x),
\label{eqn:f-xi-b}
\end{align}
where $A(x) \equiv S_1(x) \tanh x$ and $B(x) \equiv S_2(x) \tanh x$ are derived from the
following equations as
\begin{align}
& S_1(x) = \frac{4x}{\pi^4} {\rm Re} \sum_{n=0}^\infty
\frac{\ds \psi\left( 1+n-\ri{\frac{x}{\pi}} \right) - \psi\left({\frac{1}{2}}\right)}
{\ds \left(n+\frac{1}{2}\right) \left(n+\frac{1}{2}-\ri\frac{x}{\pi}\right)^2}, \cr
& S_2(x) = \frac{8}{\pi^3} {\rm Im} \sum_{n=0}^\infty
\frac{\ds \psi\left( 1+n-\ri\frac{x}{\pi} \right) - \psi\left(\frac{1}{2}\right)}
{\ds \left(n+\frac{1}{2}-\ri\frac{x}{\pi}\right)^2},
\end{align}
and $\psi$ represents the digamma function.

%%%%%%%%%%%%%%%%%%%%%%%%%%%%%%%%%%%%%%%%%%%%%%%%%%%%%%%%%%%%%%%%%%%%%%%%%%%%%%%%%%%%%%%%%%%%%%%%%%%
\section{Multipolar Scattering Anisotropy in a Magnetic Field}
%%%%%%%%%%%%%%%%%%%%%%%%%%%%%%%%%%%%%%%%%%%%%%%%%%%%%%%%%%%%%%%%%%%%%%%%%%%%%%%%%%%%%%%%%%%%%%%%%%%

It is convenient to divide $F_{\xi,\alpha}$ in eq.~(\ref{eqn:Fxi-a}) into the $\bar{h}_z$-independent
term
\begin{align}
F_{\xi} (x_0,x_h) = \frac{1}{3} \left[ f_{\xi} (x_0) + f_{\xi} (x_+) + f_{\xi} (x_-) \right]
\label{eqn:Fx0}
\end{align}
and the $\bar{h}_z$-dependent term
\begin{align}
\eta_{\xi} (x_0,x_h) = - \frac{1}{8} \left( \frac{1}{3} - \bar{h}_z^2 \right)
\left[ 2 f_{\xi} (x_0) - f_{\xi} (x_+) - f_{\xi} (x_-) \right],
\label{eqn:eta}
\end{align}
so that $F_{\xi,\alpha}$ is rewritten as
\begin{align}
& F_{\xi, x} = F_{\xi, y} = F_{\xi} (x_0,x_h) - \eta_{\xi} (x_0,x_h), \\
& F_{\xi, z} = F_{\xi} (x_0,x_h) + 2 \eta_{\xi} (x_0,x_h).
\end{align}
The $z$ component of the field $\bar{h}_z$ ($0 \le \bar{h}_z \le 1$) represents a field direction:
[110] ($\bar{h}_z = 0$), [001] ($\bar{h}_z = 1$), and [111] ($\bar{h}_z = 1 / \sqrt{3}$).
One can see that $T_{\rm c}$ has a field orientation $\bar{h}_z$ dependence, which comes from
eq.~(\ref{eqn:eta}).
For $x_h \ll 1$, it is reduced to
\begin{align}
\eta_{\xi} (x_0,x_h) = \frac{1}{8} \left( \frac{1}{3} - \bar{h}_z^2 \right) f''_{\xi} (x_0) x_h^2,
\label{eqn:eta2}
\end{align}
where $f''$ indicates the second derivative.

%%%%%%%%%%%%%%%%%%%%%%%%%%%%%%%%%%%%%%%%%%%%%%%%%%%%%%%%%%%%%%%%%%%%%%%%%%%%%%%%%%%%%%%%%%%%%%%%%%%

\end{document}